\documentclass[12pt,a4paper]{article}
\usepackage{latexsym}

\def\nn{\nonumber}
\def\K1{{\cal K}_{\bf 1}}
\def\Q20{{\cal Q}_{\bf 20'}}

\newcommand{\ie}{{\em i.e.~}}
\newcommand{\eg}{{\em e.g.~}}

\newcommand{\be}{\begin{equation}}
\newcommand{\ee}{\end{equation}}
\newcommand{\ba}{\begin{eqnarray}}
\newcommand{\ea}{\end{eqnarray}}

\begin{document}

\begin{titlepage}
\thispagestyle{empty}

\begin{flushright}
 ROM2F/2013/11 \\
\end{flushright}

\vspace{1.5cm}

\begin{center}

{\LARGE {\bf
Constraining conformal field theory \\ with  higher  spin symmetry \\ in four   dimensions  \rule{0pt}{20pt} 
}} \\
\vspace{1cm} \ {\Large Yassen~S.~Stanev  } \\
\vspace{0.6cm}  {{\it I.N.F.N.\ -\ Sezione
di Roma  Tor Vergata}} \\ {{\it Via della Ricerca  Scientifica, 1}}
\\ {{\it 00133 \ Roma, \ ITALY}}
\end{center}

\vspace{1cm}

\begin{abstract}

We analyze the constraints on the general form and the singularity structure of the correlation functions of the symmetric, traceless and conserved stress-energy tensor implied by conformal invariance and higher spin symmetry in four dimensions. In particular, we show that all these correlation functions 
will have at most double pole singularities. We then compute the 4-, 5- and 6-point functions of the stress-energy tensor and find that they are linear combinations of the three free field expressions (scalar, fermion and Maxwell field). This is a strong indication that all such theories are essentially free.

\end{abstract}


\end{titlepage}

\vfill
\newpage

\section{Introduction}
\label{sec:INTRO}

Recently there has been an impressive progress in the understanding of Conformal invariant quantum Field Theories (CFT) with Higher Spin (HS) symmetry. 
All the 3-point functions of the HS conserved currents have been found, 
first in three dimensions \cite{Giombi}, then in four \cite{GenCFT}
and in general $D$ dimensions \cite{Zhiboedov:2012bm}. All the $n$-point functions of the free HS conserved currents were derived in $D=3$   \cite{Didenko:2012tv} and $D=4$ 
dimensions  \cite{Gelfond:2013xt}. In \cite{Boulanger:2013zza} it was shown that 
introducing a single HS current always leads to an infinite tower of HS currents.

A quite different approach, particularly suited for the study of $D=4$ conformal theories, is based on the notion of Global Conformal Invariance (GCI) \cite{NT}, and makes use of bilocal and biharmonic conformal fields \cite{biharm1}. In particular, a GCI theory always contains only fields of integer scale dimension, and all the $n$-point correlation functions are rational functions. In this setup it was shown that in any GCI theory there are infinitely many HS conserved currents \cite{Rational1}. 
Moreover, if the algebra of observables is generated by  a hermitian scalar field $\Phi_2$ of scale dimension two, then all the functions can be realized in terms of free scalar fields \cite{Rational2,BNRT}.

In $D=3$ dimensions a stronger result has been proven in \cite{MZ}, 
namely that all the correlation functions of the observables  in any CFT  
with higher spin symmetry (\ie   with an infinite number of 
conserved HS currents) are equal to the ones in a free field theory. 
This can be viewed as an extension of the Coleman-Mandula theorem \cite{Colem-M} to the CFT case.

In this paper we report some progress in establishing the same result in $D=4$ dimensions. We analyze the constraints on the correlation functions of the stress-energy tensor implied by conformal invariance and HS symmetry.
An important role in our construction is played by an auxiliary scalar biharmonic field $V_2(x_1,x_2)$ which projects on the contributions of the HS conserved currents in the stress-energy tensor Operator Product Expansion (OPE)
and considerably simplifies the treatment of the correlation functions of the
 stress-energy tensor.

Let us first state our assumptions:
\begin{itemize}
\item Conformal invariant theory with all the standard properties in $D=4$ Minkowski space (existence of a stress-energy tensor, OPE, cluster decomposition, etc.). 

\item The stress-energy tensor $\Theta_{\mu \nu}(x)$ is symmetric, traceless and conserved. 

\item Unitarity  $\Leftrightarrow$ Wightman positivity of the theory. This implies that all the fields which appear in the OPE $\Theta_{\mu \nu} \Theta_{\rho \tau}$ belong to unitary representations of the conformal group.

\item Higher spin symmetry. This implies that there are infinitely many conserved HS currents \cite{Boulanger:2013zza}.  

\item Rationality. This is a rather technical assumption about the leading on the light cone part of the total contribution of all the conserved HS currents to the $n$-point functions of the stress-energy tensor spelled out in Section~3. (see Eq.(\ref{W0}) and the discussion after it). One can argue that it should hold for any CFT with HS symmetry, but we were not able to prove it in general.
To make the discussion simpler, in the rest of the paper we shall work in the GCI framework, where rationality is automatically satisfied.
\end{itemize} 
The main results of our analysis can be summarized as follows. Under the above assumptions 
\begin{itemize}
  
\item We prove that in any CFT with higher spin symmetry all the $n$-point functions of the stress-energy tensor will have at most double pole singularities in all arguments (a property typical for the free field theories).

\item We then compute the 4-, 5- and 6-point correlation functions of the  stress-energy tensor and find that they are linear combinations of the three free field expressions (scalar, fermion and Maxwell field)   Eq.(\ref{T6}). 
\end{itemize}
Our results essentially imply that any such theory, and in particular any GCI theory, is free.  

The paper is organized as follows, in Section~2 we review some properties of the stress-energy tensor in $D=4$ conformal field theory. In Section~3 we construct the auxiliary scalar biharmonic field $V_2(x_1,x_2)$ and determine the structure of the singularities of the correlation functions of the stress-energy tensor. In Section~4 we sketch the computation of the 4-, 5- and 6-
point functions of the stress-energy tensor. Finally, in Section~5 we give our conclusions and list some open problems.

\section{The stress-energy tensor $\Theta_{\mu \nu}(x)$}
\label{sec:OPE}

In this Section we shall briefly review some properties of the stress-energy tensor $\Theta_{\mu \nu}(x)$, its OPE 
and its correlation functions in four dimensional conformal field theory. 

The stress-energy tensor $\Theta_{\mu \nu}(x)$ plays a central role in any conformal invariant quantum field theory since it generates the conformal transformations. 
It is symmetric, traceless and conserved 
\be
\Theta_{\mu \nu}(x) = \Theta_{\nu \mu}(x) \ , \quad
{\Theta^\mu}_\mu(x) = 0 \ , \quad 
\partial_{\mu} \Theta^{\mu \nu}(x) = 0 \ ,   
\label{Theta}
\ee
has scale dimension $\Delta_\Theta=4$ and under special conformal transformations transforms as
\ba 
\left[ C_{\alpha} , \Theta_{\mu \nu}(x) \right] &=& 
(2 x_\alpha (x.\partial_x + 4) -x^2 \partial_{x^\alpha}) \Theta_{\mu \nu}(x)
+2 \, \eta_{\mu \alpha}  x^{\rho} \Theta_{\rho \nu}(x) \nonumber \\ 
&+& 2 \, \eta_{\nu \alpha}  x^{\rho} \Theta_{\mu \rho}(x)
-2 \, x_{\mu} \Theta_{\alpha \nu}(x)
-2 \, x_{\nu} \Theta_{\mu \alpha}(x) \ . 
\label{Tconf}
\ea

In a theory with higher spin symmetry, there are also (infinitely many) 
higher spin conserved currents, namely symmetric traceless conserved tensors $J_r^{(\mu_1 \dots \mu_r)}(x)$ of rank $r$ and scale dimension $\Delta_r=r+2$, transforming in the Lorentz representation $(r/2,r/2)$.  For a symmetric tensor, the quantity $\Delta_r-r$ is called the twist, so all the conserved currents $J_r(x)$ have twist two. The $r=0$ case is special, it corresponds to a scalar field $\Phi_2$ of scale dimension equal to two which does not obey a conservation condition, but it shares many properties with the family of conserved currents. The $r=1$ field is the usual
dimension three conserved current $J_{\mu}(x)$, while the  stress-energy tensor $\Theta_{\mu \nu}(x)$ is (one of) the $r=2$ currents.

In this paper we shall consider the $n$-point truncated Wightman functions, vacuum expectation values of ordinary (not time-ordered) products of $n$ fields, of the stress-energy tensor  $\Theta_{\mu \nu}(x)$
\be
{\cal G}_n(x_1,x_2,\dots,x_n) = \langle \Theta^{\mu_1 \nu_1}(x_1) \Theta^{\mu_2 \nu_2}(x_2) \dots \Theta^{\mu_{n} \nu_{n}}(x_{n}) \rangle \vert_{\rm truncated}
\, ,
\label{defGn}
\ee
where, as usual, truncated means that the terms which are products of lower point functions are subtracted . For example,  from the 4-point function we subtract the three possible products of two 2-point functions.
Because of the cluster decomposition property, the truncated functions vanish 
asymptotically 
whenever two subsets of the arguments are confined in two distinct regions which are at large space-like distance. Local commutativity implies that the function  ${\cal G}_n$ is totally symmetric under the permutation of any two arguments (accompanied by the permutation of the respective Lorentz indices).
The properties of the stress-energy tensor $\Theta_{\mu \nu}(x)$ imply that ${\cal G}_n$ is a parity even, symmetric and traceless in each pair of indices  $(\mu_i,\nu_i)$ and conserved in all its arguments function. The conformal transformation law for $\Theta_{\mu \nu}(x)$, Eq.(\ref{Tconf}), determines also the transformation properties of ${\cal G}_n$ under special conformal transformations. 

It follows that ${\cal G}_n$ (very schematically)  will have the form
\be
 {\cal G}_n \ = \ P(\{x_{ij}^2\}) \times  \sum_a Q_a(\{R(jk),L(ijk)\}) \, f_a(\{s_{i j k \ell}\}) \ ,
\label{generalnp}
\ee
where $x_{ij}=x_i-x_j$, the prefactor $P$ takes care of the scale dimension
of the function,   
while $Q_a$ are appropriate polynomials of the primitive 2- and 3-point covariants $R(jk)$ and $L(ijk)$  
\ba 
 R(jk) &=&  \frac{1}{x_{jk}^2}\left( {\eta^{\mu_j \mu_k} - 2 \ \frac{ x_{jk}^{\mu_j} \ x_{jk}^{\mu_k}}{x_{jk}^2}}\right) \ , 
\nonumber \\
L(ijk) &=&   \frac{x_{ij}^{\mu_i}}{x_{ij}^2} - \frac{x_{ik}^{\mu_i}}{x_{ik}^2}
\ .  
\label{LR}
\ea
Finally, $f_a(\{s_{i j k \ell}\})$ are functions of the conformal invariant cross-ratios
\be 
s_{i j k \ell} \ = \ \frac{x_{ij}^2 x_{k \ell}^2} {x_{ik}^2 x_{j \ell}^2} \ , 
\label{crossR}
\ee 
which are present only for 4- and higher- point functions.
Note that the representation Eq.(\ref{generalnp}) is highly non-unique.
On the one hand both $L(ijk)$ and $s_{i j k \ell}$ are non independent,
for example $L(ij\ell)=L(ijk)+L(ik\ell)$. On the other hand for the 4- and higher- point functions the separation in $P$  and $f_a$ is also a matter of choice.

Imposing the symmetries and the conservation leads to conditions for
$Q_a$ and the products $P f_a$. In particular,  
the 2-point function is unique (up to normalization), while
the space of 3-point functions is three dimensional \cite{Me88}. A convenient choice of basis in this case is given by the 3-point 
functions of the stress energy tensor in the theory of a free scalar $\varphi$, a free fermion $\psi$ and a free Maxwell field $F_{\mu \nu}$. 
For $n \geq 4$  some of the functions $f_a$ remain undetermined.
Additional constraints on the leading short distance singularities of the 
functions Eq.(\ref{generalnp}) follow from the OPE.

The OPE of two stress-energy tensors is rather complicated. Here we shall briefly review only some relevant for our discussion properties. The general form of the OPE is
\be
 :\Theta_{\mu \nu}(x_1) \, \Theta_{\rho \tau}(x_2): \ = \   \sum_{R}
\  {\cal C}^{R}_{\mu \nu \rho \tau} (z,\partial) \ {\cal O}_R(x_2) \, ,
\label{TTOPE}
\ee
where $z=x_1-x_2$, the Lorentz indices of ${\cal O}_R$ contracted with 
${\cal C}^R$ are suppressed, and the normal product $ : \ : $ as usual denotes the subtraction of the 2-point function of the stress-energy tensor.
The label $R =(j_1,j_2;\Delta)$ defines the representation of the conformal group  to which belongs the field ${\cal O}_R$,
where $(j_1,j_2)$ label the representation of the Lorentz group,
while $\Delta$ is the (integer in the case of interest) scale dimension.
In principle, any of the representations of the Lorentz group $(j,j+k)\oplus(j+k,j)$ with $j$ arbitrary non-negative (half)integer and integer $k=0,\dots,4$ may appear. 

In general in the right hand side of the OPE there can be both parity even and parity odd fields ${\cal O}_R$.  Since the left hand side is parity even,  
the parity odd contributions will always be multiplied by the totally antisymmetric tensor $\epsilon_{\mu \rho \alpha \beta}$.

Each of the coefficients ${\cal C}^{R}$ is a linear combination of terms, which in turn can be expanded as formal double power series in $(z \partial)$ and 
$(z^2 \Box)$. Hence we may write (suppressing all Lorentz indices) 
\be
{\cal C}^{R} = \sum_{\ell} {{\cal M}^{(R,\ell)}(z,\partial) \over
(z^2)^{N(R,\ell)}} \sum_{k,n=0}^{\infty} \alpha_{k,n}^{(R,\ell)} (z \partial)^k (z^2 \Box)^n \ , 
\label{Cseries}
\ee
where the sum in $\ell$ is over all possible monomials ${\cal M}^{(R,\ell)}$ in $z$ and $\partial$, with the same Lorentz structure as ${\cal C}^{R}$, and $N(R,\ell)$ are integer. Imposing conformal invariance allows to express all the coefficients $\alpha_{k,n}^{(R,\ell)}$ in terms of some of the 
$\alpha_{0,0}^{(R,\ell)}$. The explicit solution for the stress-energy tensor  contribution, ${\cal O}_R=\Theta_{\mu \nu}$, was found in \cite{Me88}, where also the 54 relevant monomials ${\cal M}^{(R,\ell)}(z,\partial)$  were written and it was shown that there are exactly three independent conformal structures.

Fortunately we shall need something much simpler. Let us denote the 
scale dimension of ${\cal M}^{(R,\ell)}(z,\partial)$ by  $M(R,\ell)$.
Scale invariance of Eqs.(\ref{TTOPE},\ref{Cseries}) implies the equation 
\be
2 \, \Delta_{\Theta} = 8 = 2 \, N(R,\ell) - M(R,\ell) + \Delta_R \ , 
\label{scale}
\ee
for each $R$ and each $\ell$, which leads to an upper bound for each $N(R,\ell)$.
Indeed only unitary conformal representations may appear in the OPE
of two stress-energy tensors
and unitarity gives lower bounds on the scale dimensions of the fields \cite{Mack}. In particular, if a field belongs to the representation $R=(j_1,j_2;\Delta_R)$, its scale dimension $\Delta_R$  has to satisfy 
\ba 
\Delta_R &\geq& j_1+j_2+1 \quad {\rm if} \quad j_1 j_2=0 \ , \nn \\
\Delta_R &\geq& j_1+j_2+2 \quad {\rm if} \quad j_1 j_2 \neq 0 \ .
\label{unitary}
\ea  
Combining these inequalities with Eq.(\ref{scale}), 
a very tedious, but straightforward case by case analysis shows that    
 the coefficient functions in the OPE 
\be
a^\mu a^\nu  b^\rho b^\tau {\cal C}^R_{\mu \nu \rho \tau} (z,\partial)\, , 
\label{OPE2}
\ee
where we introduced the auxiliary vectors $a$ and $b$ to make the symmetry manifest,  should respect for any $R$ the following pole bounds   
\ba
{1 \over {(z^2)}^N} &{\rm is \ forbidden \ for \ any } \ N \geq 6 & \, , \nonumber \\
{1 \over {(z^2)}^5} &{\rm is \ forbidden \ unless \ multiplied \ by}&  \  (az)^2(bz)^2 \, , \nonumber \\
{1 \over {(z^2)}^4} &{\rm is \ forbidden \ unless \ multiplied \ by}&  \ (az) \ {\rm or } \ (bz) \, ,
\nonumber \\
{(ab) \over {(z^2)}^4} &{\rm is \ forbidden \ unless \ multiplied \ by}&  \ (az) (bz)
\, . 
\label{polebounds}
\ea
The poles of order less than four are not constrained.
The meaning of the last two of the above conditions is as follows. A pole of order four can appear only if the monomial which multiplies it contains as a factor either $(az)$ or $(bz)$. If the monomial has a factor $(ab)$, then it should contain also both factors $(az)$ and $(bz)$. 
Note that these are the most conservative estimates, valid for any representation $R$ and any parity of the fields. They can be refined by 
restricting to only some representations or by  
considering separately the contributions from the parity even and parity odd fields. Moreover, whether certain pole structure will effectively appear is model dependent. For example the functions 
of the stress-energy tensor in the theory of free scalar field have maximal poles of order five,  in the theory of free fermion field have maximal poles of order four, while  in the theory of free Maxwell field have maximal poles of order three. The pole bounds for the coefficients in the OPE are in one to one correspondence with the leading light-cone and short distance singularities of the (truncated) correlation functions of the stress energy tensor $\Theta_{\mu \nu}(x)$,  Eq.(\ref{defGn}), but imposing the pole bounds 
still not sufficient to determine completely the functions in Eq.(\ref{generalnp}). 
Additional conditions may follow from Wightman positivity, which for the case of 4-point functions boils down to the condition that all the coefficients in the expansion of ${\cal G}_n$ in conformal partial waves should be positive, but this is a very complicated non-linear problem and 
requires the knowledge of the conformal partial waves. Thus we shall not discuss 
Wightman positivity. Instead, we shall consider only the class of rational functions. If the prefactor $P$ is properly chosen, this implies that all the functions $f_a$ in Eq.(\ref{generalnp}) will be polynomials in  $s_{i j k \ell}$, of maximal degree not higher than five due to the pole bounds in Eqs.(\ref{polebounds}). This reduces the problem to a finite dimensional one, since instead of dealing with unknown functions, we have to deal with (a large number of) unknown coefficients of the polynomials $f_a$.  
Still the problem remains very difficult. This is partly due to the large number of Lorentz structures in the game, as well as to the linear dependence of the 
covariants $L(ijk)$, but we think that the main reason is that in the representation in terms of primitive conformal covariants Eq.(\ref{generalnp})  some essential features of the $n$-point functions of the stress energy tensor are not manifest. We shall return to this in the next Section.

\section{Biharmonic field construction}
\label{sec:Biharm}

In this Section we shall develop a formalism for the computation of the correlation functions of the stress-energy tensor based on the systematic 
use of (biharmonic) bi-fields.

Given the OPE of  two stress-energy tensors $\Theta_{\mu \nu}$, Eq.(\ref{TTOPE}), one can construct the auxiliary  
bi-field $V(x_1,x_2)$  
\ba
V(x_1,x_2) &=&  x_{12}^2  \, \left( x_{12}^\mu x_{12}^\nu x_{12}^\rho x_{12}^\tau 
- x_{12}^2 \, x_{12}^\mu x_{12}^\rho \, \eta^{\nu \tau} + {x_{12}^4 \over 4}  \, \eta^{\mu \rho} \, \eta^{\nu \tau} 
 \right)  \nonumber  \\
&\times& \, :\Theta_{\mu \nu}(x_1) \, \Theta_{\rho \tau}(x_2):  \ , 
\label{defV}
\ea
where $x_{12}=x_1-x_2$ and the normal product $ : \ : $ denotes the subtraction of 
the 2-point function of the stress-energy tensor.

The bi-field  $V(x_1,x_2)$ has a number of interesting properties which we shall now list and comment.

The transformation law for $\Theta_{\mu \nu}$, Eq.(\ref{Tconf}), implies that 
under special conformal transformations  $V(x_1,x_2)$ transforms as a scalar conformal bi-field of weights $(1,1)$ in $x_1$ and $ x_2$ respectively.

Since the OPE $\Theta_{\mu \nu}(x_1) \, \Theta_{\rho \tau}(x_2)$  is symmetric (up to contact on the light cone terms, which due to the overall $x_{12}^2$ factor in Eq.(\ref{defV}) will not contribute to $V$) under the simultaneous exchange of $(x_1,\mu,\nu)$ and $(x_2,\rho,\tau)$, it follows that $V(x_1,x_2)$ is symmetric under the exchange of
$x_1$ and $x_2$. Hence it receives contributions from only the even rank symmetric tensors in the OPE $\Theta_{\mu \nu}(x_1) \, \Theta_{\rho \tau}(x_2)$.

The bi-field $V(x_1,x_2)$ is finite in the limit $x_{12} \rightarrow 0$.
This follows by inspection of the leading singular terms in the OPE $\Theta_{\mu \nu}(x_1) \, \Theta_{\rho \tau}(x_2)$.
The lowest dimensional contribution in $V(x_1,x_2)$  is the scalar field $\Phi_2$ of scale dimension two  (when present). This $\Phi_2$ contribution is in one-to-one correspondence with the non-zero limit of $V(x_1,x_2)$ for coincident arguments $x_2 = x_1$. The case of a GCI theory 
generated by the scalar field $\Phi_2$ has been analyzed in detail in \cite{Rational1}, \cite{BNRT} and it has been shown that it reduces to a 
theory of free scalars. What is relevant for our purposes is that all the 
functions of the stress energy tensor  are also given by the expressions in the theory of a free scalar field. Since we are looking for non-trivial solutions for the functions of the stress-energy tensor, without loss of generality we may 
assume that there is no $\Phi_2$ contribution in $V(x_1,x_2)$, or equivalently that $V(x_1,x_2)$ is zero in the limit $x_{12} \rightarrow 0$. Hence, for the rest of this Section we shall assume\footnote{In the next Section we shall explain how one can treat the case when there is a $\Phi_2$ contribution.}
\be
V(x_1,x_1) \ = \ 0 \, . 
\label{Vzero}
\ee

In the light-cone limit, when $x_{12}^2 \rightarrow 0$, only the even rank twist=2 conserved currents $J_r$ contribute to $V(x_1,x_2)$. This again follows by inspection of the singularities of the OPE  $\Theta_{\mu \nu}(x_1) \, \Theta_{\rho \tau}(x_2)$.

Even if several different rank two symmetric traceless conserved
tensors contribute to the $\Theta \Theta$ OPE, only the stress-energy tensor
$\Theta_{\mu \nu}$ contributes to  $V(x_1,x_2)$. 
To prove this let us assume that in  the $\Theta \Theta$ OPE there is also another rank two symmetric traceless and conserved tensor $T_{\mu \nu}$, orthogonal to $\Theta_{\mu \nu}$ in the sense that their 2-point function vanishes $\langle \Theta_{\mu \nu} T_{\rho \tau} \rangle = 0$. We recall that there are exactly three independent conformal invariant 3-point functions of three (non necessarily equal) rank two symmetric traceless and conserved tensors, namely the 3-point 
functions of the stress energy tensor in the theory of a free scalar $\varphi$, a free fermion $\psi$ and a free Maxwell field $F_{\mu \nu}$. Thus any 3-point function $\langle \Theta \, \Theta \, T \rangle$
can be written as
\be 
\langle \Theta_{\mu \nu}(x_1) \, \Theta_{\rho \tau}(x_2) \, T_{\alpha \beta}(x_3)  \rangle = c_{\varphi}\langle\Theta\Theta\Theta\rangle_{\varphi}
+c_{\psi}\langle\Theta\Theta\Theta\rangle_{\psi}
+c_{F}\langle\Theta\Theta\Theta\rangle_{F} \, .
\label{ThThT}
\ee
The generators of the conformal transformations can be expressed as integrals of the stress-energy tensor.
In particular, the translation generators are 
\be
P_{\mu} = \int d^3x \, \Theta_{0 \mu }(x) \, ,
\label{translation}
\ee
hence we have 
\be
 \left[ \int d^3x_1 \, \Theta_{0 \mu }(x_1), \Theta_{\rho \tau}(x_2) \right] 
= \left[ P_{\mu}, \Theta_{\rho \tau}(x_2) \right]  
= -{\rm i} \partial_{\mu} \Theta_{\rho \tau}(x_2) \, .  
\label{TTComm}
\ee
Since $\Theta_{\mu \nu}$ and $T_{\mu \nu}$ are orthogonal, it follows that
\be 
\langle \left[ \int d^3x_1 \, \Theta_{0 \mu }(x_1), \Theta_{\rho \tau}(x_2) \right]  \, T_{\alpha \beta}(x_3)  \rangle \ = \
 -{\rm i} 
\partial_{\mu} \langle \Theta_{\rho \tau}  \, T_{\alpha \beta}   \rangle  \ = \ 0 \, ,
\label{Theta21}
\ee
which implies a relation for the coefficients $ c_{\varphi}, c_{\psi}$ and  $c_{F}$  in Eq.(\ref{ThThT}). It is straightforward to verify that this relation in turn implies 
\be
\langle V(x_1,x_2) \, T_{\alpha \beta}(x_3)  \rangle \ = \ 0 \, , 
\label{VT}
\ee
where $\langle V(x_1,x_2) \, T_{\alpha \beta}(x_3)  \rangle$  is
obtained from the 3-point function in Eq.(\ref{ThThT}) by applying  Eq.(\ref{defV}).
Thus the bi-field $ V(x_1,x_2)$, defined in Eq.(\ref{defV}), 
is orthogonal to any field $T_{\mu \nu}$ orthogonal to the stress-energy tensor 
$\Theta_{\mu \nu}$.

Although $V(x_1,x_2)$ in general receives contributions also from the higher twist  operators, the scalar field $\Phi_4$ of scale dimension four is  projected out. This follows by explicitly computing the unique conformal invariant 3-point function $\langle \Theta_{\mu \nu}(x_1) \, \Theta_{\rho \tau}(x_2) \Phi_4(x_3) \rangle$, which implies  $\langle V(x_1,x_2) \Phi_4(x_3) \rangle$ = 0. Since by assumption the scalar field $\Phi_2$ is also absent, the stress-energy tensor $\Theta_{\mu \nu}$ is the leading (for small $x_{12}$) contribution in $V(x_1,x_2)$ and can be expressed  as 
\be
\Theta^{\mu \nu}(x_1) = 
\left( \partial_{x_{12}}^{\mu} \partial_{x_{12}}^{\nu} 
- { \eta^{\mu \nu} \over 4} \, \Box_{x_{12}} \right) V(x_1,x_2)\vert_{x_{12}=0}  \, . 
\label{ThetafromV}
\ee

To summarize, Eqs.(\ref{defV}),(\ref{ThetafromV}) define a fusion procedure
$ :\Theta \Theta: \rightarrow V \rightarrow \Theta $. This fusion relates the 
$n$- and $n+1$-point functions of $\Theta$, and is much easier 
to impose than the integral relation in Eq.(\ref{TTComm}).
It also allows to obtain from the  $2n$-point function of $\Theta$ the $2n$-point functions of $n$ scalar bi-fields $V$, 
which in turn can be reduced to the $n$-point function of $\Theta$. 
As we shall argue below, the structure of the intermediate expression for the scalar bi-fields $V$ is simpler than the expressions for $\Theta$. 

We continue our study of the properties of the bi-field $V(x_1,x_2)$.
Let us first note that 
all the 3-point functions of 
$V(x_1,x_2)$ and a HS conserved current $J_r(x_3)$ of rank $r$  satisfy the biharmonicity condition
\be
\Box_{x_1} \langle V(x_1,x_2) \, J_r(x_3) \rangle \ = \ 
\Box_{x_2} \langle V(x_1,x_2) \, J_r(x_3) \rangle \ = \ 0 \, 
\label{biharmonic1}
\ee
for any $r$.
This follows by a direct calculation, first extracting from the generating function for all the 3-point functions of conserved currents \cite{GenCFT} the  symmetric in $x_1 \leftrightarrow x_2$ 3-point functions of two 
rank two conserved currents and a rank $r$ conserved current, then computing 
$\langle V \ J_r \rangle$.  
If we define the restriction of $V$ to only the twist=2 contributions
in the stress-energy tensor OPE 
\be
V_2(x_1,x_2) = V(x_1,x_2) \vert_{twist=2} = \sum_r {\cal C}_r(x_{12},\partial_{x_2}) \,  J_r(x_2) \, ,
\label{V2}
\ee
Eqs.(\ref{biharmonic1}) can be compactly rewritten as 
\be
\Box_{x_1} \langle V(x_1,x_2) \, V_2(x_3,x_4)) \rangle \ = \ 
\Box_{x_2} \langle V(x_1,x_2) \, V_2(x_3,x_4)) \rangle \ = \ 0 \, ,
\label{biharmonic2}
\ee
which in turn imply (because of the orthogonality of the different 
twist contributions)
\be
\Box_{x_1} \langle V_2(x_1,x_2) \, V_2(x_3,x_4)) \rangle \ = \ 
\Box_{x_2} \langle V_2(x_1,x_2) \, V_2(x_3,x_4)) \rangle \ = \ 0 \, .
\label{biharmonic3}
\ee
Wightman positivity then implies that $V_2$ is biharmonic as operator identity
\be
\Box_{x_1} V_2(x_1,x_2)  \ = \ 
\Box_{x_2}  V_2(x_1,x_2)  \ = \ 0 \, .
\label{biharmonic4}
\ee
Note that without loss of generality one can replace in Eq.(\ref{ThetafromV}) $V(x_1,x_2)$ with $V_2(x_1,x_2)$.

Let us stress that this construction applies to any CFT. The bi-field $V$ and its harmonic part $V_2$ can be defined also in theories without HS symmetry\footnote{In this case $V_2$ will contain only the stress-energy tensor.}, and even in theories like ${\cal N}=4$ SYM where some of the fields in the  $\Theta \Theta$ OPE have anomalous dimensions\footnote{In this case Eq.(\ref{ThetafromV}) should be modified by subtracting from the r.h.s. all the fields with scale dimension $\Delta < 4$.}.
Thus, one possible way of constructing all the possible conformal invariant $n$-point functions of the stress-energy tensor $\Theta_{\mu \nu}$ in any CFT is to classify all the $2n$ point functions of $V_2$, harmonic in all the arguments and then use  Eq.(\ref{ThetafromV}) to derive the respective functions of the stress-energy tensor. 

Computing the general harmonic in all its arguments  function is an extremely difficult task. 
Hence, to proceed, we shall make a simplifying assumption. Namely, motivated by GCI\footnote{As already mentioned, any GCI theory has HS symmetry.}, we shall assume rationality of the functions of the stress-energy tensor ${\cal G}_n$.    
Even in this case at first sight there is a problem.
To spell it out let us first introduce some notation. We shall denote the $2n$-point 
Wightman function of $n$  bi-fields $V$  by ${\cal W}(2n)$
\be 
{\cal W}(2n) = \langle V(x_1,x_2) V(x_3,x_4) \dots V(x_{2n-1},x_{2n}) \rangle \, .
\label{W}
\ee
 We shall denote the $2n$-point 
Wightman function of $n$  biharmonic fields $V_2$  by 
${\cal W}_2(2n)$
\be 
{\cal W}_2(2n) = \langle V_2(x_1,x_2) V_2(x_3,x_4) \dots V_2(x_{2n-1},x_{2n}) \rangle \, .
\label{W2}
\ee
Both functions are scalar conformal invariant functions of conformal weights in all the 
arguments equal to one. Hence they both will depend only on\footnote{We shall systematically omit the $+{\rm i} 0  x_{jk}^0$ prescription.}
\be
Z_{jk} = x_{jk}^2 \ ,
\label{Z}
\ee 
and will be non-singular for $x_2=x_1$, $x_4=x_3$, \dots , $x_{2n}=x_{2n-1}$.
However there is an important difference. 
Since  ${\cal W}(2n)$ is obtained from the rational function  ${\cal G}_{2n}$
by multiplication with $x_{ij}$ and contracting the Lorentz indices,  
 it is also a rational function 
and will have the general form
\be 
 \sum_{m_{jk}} \,  C_{m_{jk}} \, \prod_{j<k}^{2n} Z_{jk}^{m_{jk}}  \, , 
\label{Wrat}
\ee
where the sum is over all configurations of integer powers $m_{jk} = m_{kj}$
satisfying $\sum_{j\neq k}^{2n} m_{jk} = -1$ for any $k$. Note that the last equation implies that for any fixed $k$ at least one of the integers $m_{jk}$ will be negative, hence each term in (\ref{Wrat}) will be singular when $x_k$
approaches some of the other points.  

On the contrary, the function ${\cal W}_2(2n)$, which is harmonic in all its arguments, is not related in a simple way to ${\cal G}_{2n}$, and apriori is not a rational function. 

The problem can be avoided by the observation that the restrictions of the functions ${\cal W}(2n)$ and ${\cal W}_2(2n)$ for
$Z_{12}=0$, $Z_{34}=0$, \dots, $Z_{2n-1 \, 2n}=0$ coincide, since in this limit one projects out all the contributions of the higher twist fields (as well as some of the higher derivatives of the twist=2 fields) 
\ba
{\cal W}_0(2n) &=& {\cal W}(2n)\vert_{Z_{12}=0, Z_{34}=0, \dots, Z_{2n-1 \, 2n}=0} \nonumber \\
&=& {\cal W}_2(2n)\vert_{Z_{12}=0, Z_{34}=0, \dots, Z_{2n-1 \, 2n}=0} \ .
\label{W0}
\ea
Hence, on the one hand ${\cal W}_0(2n)$ is a rational function and will have the form  
(\ref{Wrat}), on the other hand it can be completed to a harmonic in all 
its arguments function. As already stressed in the Introduction, the existence of ${\cal W}_0(2n)$ with these two properties is sufficient for our construction to hold. Thus we can replace the requirement that all the $n$-point correlation functions of the stress-energy tensor ${\cal G}_n$ are rational functions by 
the requirement that all the functions ${\cal W}_0(2n)$ obtained 
from ${\cal G}_{2n}$ are rational. Since ${\cal W}_0(2n)$ captures only (the leading on the light-cone part of) the contributions of the conserved currents, we expect this to hold in any CFT with HS symmetry.   
Note that, while any rational function can be 
completed to a harmonic in one of its arguments function \cite{BargTod}, the 
requirement that there exists a completion harmonic in all arguments is highly non-trivial. 

The partial restrictions, \eg when only $Z_{12}=0$, will also be rational and of the form (\ref{Wrat})
\be
 {\cal F}(2n ;\, 1,2)  =  {\cal W}(2n)\vert_{Z_{12}=0} =  \sum_{m_{jk}} \,  C_{m_{jk}} \, \prod_{j<k}^{2n} Z_{jk}^{m_{jk}} \, ,
\label{Wrest}
\ee
and can be completed to a biharmonic in $x_1$ and $x_2$ function, namely
the $2n$-point Wightman function 
$ \langle V_2(x_1,x_2) V(x_3,x_4) \dots V(x_{2n-1},x_{2n}) \rangle$.
The conditions imposed by biharmonicity have been studied in detail in \cite{biharm1}. In particular, it has been shown that 
the function ${\cal F}(2n ;\, 1,2)$ defined in Eq.(\ref{Wrest}) can be completed to a biharmonic (in $x_1$ and $x_2$) function if and only if in each monomial in the r.h.s of Eq.(\ref{Wrest}) at most two of the integers $m_{1i}$ are negative and similarly for  $m_{2i}$.
Moreover, if $p$ and $q$ are such that $m_{1p}<0$ and $m_{1q}<0$, then all the coefficients $m_{2j}$ for $j \neq p,q$ will be non-negative. Putting everything together, the general singularity structure of each monomial in the r.h.s of Eq.(\ref{Wrest}) will be 
\be
{P(\{Z_{1j}\},\{Z_{2k}\}) \over {(Z_{1p})}^{\alpha_p} {(Z_{1q})}^{\alpha_q} {(Z_{2p})}^{\beta_p} {(Z_{2q})}^{\beta_q}} \times R \, , 
\label{doublepole}
\ee
where $\alpha_{p,q}=-m_{1p,q}$ and $\beta_{p,q}=-m_{2p,q}$ are non-negative integers,  
$P$ is a polynomial of its arguments,  $j,k \neq p,q$, and $R$ denotes a factor independent of $x_1$ and $x_2$. Following the notation introduced in \cite{biharm1} we shall call the structure in (\ref{doublepole}) "double pole"
and will refer to the special case when one of the two $\alpha_p$, $\alpha_q$ as well as one of the two $\beta_p$, $\beta_q$ is zero as "single pole" singularity structure.  
The importance of the single pole case is due to the observation that the biharmonic completion of a rational single pole function is again rational single pole \cite{biharm1}. 

To summarize, the function  ${\cal F}(2n ;\, 1,2)$, defined in Eq.(\ref{Wrest}),
will have at most a double pole (in $x_1$ and $x_2$) singularity structure.
Note that the singularity structure in the other $2n-2$ arguments of  ${\cal F}(2n ;\, 1,2)$ is not constrained. 

We can define also  the partial restrictions  when only $Z_{2k-1 \, 2k}=0$,  for any $k=1, \dots, n$
\be
 {\cal F}(2n ;\, 2k-1,2k)  =  {\cal W}(2n)\vert_{Z_{2k-1 \, 2k}=0} \ ,
\label{Wrest2}
\ee
and repeat the above arguments (with the obvious substitution of $1,2$ with $2k-1,2k$ in all the formulae). It follows that the function  ${\cal F}(2n ;\,  2k-1,2k)$ will have at most a  double pole (in $x_{2k-1}$ and $x_{2k}$) singularity structure.  

But then it immediately follows that the function ${\cal W}_0(2n)$, defined in Eq.(\ref{W0}), will have at most double pole singularities in all its arguments  $x_{1}, \dots, x_{2n}$. Indeed, since ${\cal W}_0(2n)$ can be obtained by restriction from any of the functions ${\cal F}(2n ;\, 2k-1,2k)$
it contains the subset of monomials present in all these functions,    
hence cannot have singularities which are not present in all ${\cal F}(2n ;\, 2k-1,2k)$ simultaneously.  

As already mentioned, given the function ${\cal W}(2n)$
one can use Eq.(\ref{ThetafromV}) to compute the $n$-point function of 
the stress-energy tensor. Note that the difference 
${\cal W}(2n)-{\cal W}_0(2n)$ by construction contains positive integer powers 
of $x_{12}^2, \ x_{34}^2, \dots,  x_{2n-1 \, 2n}^2$ and hence is annihilated either by the differential operator in the r.h.s. of Eq.(\ref{ThetafromV}), or in the limit $x_{12}=0$, $x_{34}=0$, etc. It then follows that given only the function ${\cal W}_0(2n)$ we can express the $n$-point function of the stress-energy tensor defined in Eq.(\ref{generalnp}) as
\be
{\cal G}_n(x_1,x_3,\dots,x_{2n-1}) 
 \  = \ \prod_{k=1}^{n}{\cal D}_{2k-1} {\cal W}_0(2n) \vert_{\{x_{2k} = x_{2k-1}\}}
\, ,
\label{TfromW0}
\ee
where the differential operators ${\cal D}_{2k-1}$ are obtained from
\be 
{\cal D}_{1}  = \left( \partial_{x_{12}}^{\mu_{1}} \partial_{x_{12}}^{\nu_{1}} 
- { \eta^{\mu_{1} \nu_{1}} \over 4} \, \Box_{x_{12}} \right) \,  , 
\label{CalD1}
\ee
by the substitution $1 \rightarrow 2k-1$, $2 \rightarrow 2k$.

This simple equation has a remarkable consequence. 
Since  ${\cal W}_0(2n)$ has at most double pole 
singularities (see (\ref{doublepole})), and since the differential operators ${\cal D}_j$ cannot create new singularities, Eq.(\ref{TfromW0}) implies that all the $n$-point functions of the stress-energy tensor $\Theta_{\mu \nu}$ will have also at most double pole singularities. In particular from the term  (\ref{doublepole})  one obtains
\be
{P(x_{1j}) \over {(Z_{1p})}^{m_p} {(Z_{1q})}^{m_q} } \times R \, , 
\label{tdoublepole}
\ee
where $Z_{ij}$ are defined in Eq.(\ref{Z}), $P$ is a polynomial, and $R$ is the $x_1$-independent part.
We shall denote this again as the "double pole" property, but one should distinguish between double poles in functions of ordinary fields Eq.(\ref{tdoublepole}) and 
double poles in functions of bi-fields Eq.(\ref{doublepole}). This property is trivially satisfied for the free field theories. 
Indeed for free fields, the stress-energy tensor is 
a bilinear combination of the fundamental fields and all its functions can be expressed in terms of the 2-point functions of the fundamental fields by Wick theorem, leading to double pole structure. In the usual way of writing the
conformal invariant $n$-point functions (cf. Eq.(\ref{generalnp})) the  double pole property is very
obscure (even for the free field theories), since when expressing the function in terms of the primitive conformal covariants $L$ and $R$, defined in  Eq.(\ref{LR}), one introduces fake poles, which cancel between different terms.

Thus, all the $n$-point functions of the stress-energy tensor ${\cal G}_{n}$ will have the double pole property. This in turn implies that ${\cal W}_0(2n)$, defined in Eq.(\ref{W0}), will have single pole singularity structure in all its arguments  $x_{1}, \dots, x_{2n}$. 
To prove this let us start from the $2n$-point function of the stress-energy tensor ${\cal G}_{2n}(x_{1},x_2 \dots,  x_{2n-1}, x_{2n})$, and for simplicity consider only the $x_1$ and $x_2$ dependence of the denominator. 
Because of the  double pole property ${\cal G}_{2n}$ may contain only two 
types of terms, namely
\be 
{1 \over  {(Z_{1i})}^{m_i} {(Z_{1j})}^{m_j} {(Z_{2k})}^{m_k} {(Z_{2\ell})}^{m_{\ell}} } \qquad {\rm and} \qquad    
{1 \over  {(Z_{1i})}^{m_i} {(Z_{12})}^{m_{12}} {(Z_{2k})}^{m_k} }
\, , 
\label{G2nsing}
\ee
for some  $i,j,k,\ell \neq 1,2$.
Since the bi-field $V(x_1,x_2)$ in Eq.(\ref{defV}) contains an overall $x_{12}^2 (= Z_{12})$ factor, and ${\cal W}_0(2n)$ is obtained from  ${\cal W}(2n)$, defined in Eq.(\ref{W}), by setting $x_{12}^2=0$,  it is clear 
that only the latter terms in (\ref{G2nsing}) may contribute to ${\cal W}_0(2n)$. Note also that since $V(x_1,x_2)$ is by construction finite for $x_{12}=0$, the $Z_{12}$ poles in these terms will be compensated by the explicit $x_{12}$ dependence of $V$. Hence the function ${\cal W}_0(2n)$ will have only single poles in $x_1$ and $x_2$.
The same argument can be repeated for $x_3$ and $x_4$, $x_5$ and $x_6$, etc.,  proving that ${\cal W}_0(2n)$ will have 
only single pole singularities in all its arguments  $x_{1}, \dots, x_{2n}$. 
   
Putting everything together we have proven that 
\begin{itemize}
\item All the $n$-point functions of the stress-energy tensor ${\cal G}_{n}$,
defined in Eq.(\ref{generalnp}),  will have the  double pole property.
\item All the $n$-point functions of the stress-energy tensor
${\cal G}_{n}$ will satisfy Eq.(\ref{TfromW0}), where ${\cal W}_0(2n)$,
defined in Eq.(\ref{W0}) as a restriction of a $2n$-point function ${\cal G}_{2n}$,  is a single pole rational function  which has a harmonic completion in all its $2n$ arguments.
\end{itemize}

This is the main result of this Section, so let us briefly comment on it.
The first statement is very strong, since it essentially states that even if there are non-trivial correlation functions of the stress-energy tensor their singularity structure will be of the "free field type". In particular, it excluded all the candidates for a non-trivial 4-point function of the stress-energy tensor we have found in a parallel direct search by the method briefly described in the previous Section.
On the other hand, the second statement may look rather technical, but has a very simple and intuitive physical meaning. Since the $n$-point correlation functions of the stress-energy tensor are a coupled system, related by Eqs.(\ref{TTComm}), or equivalently by Eqs.(\ref{defV},\ref{ThetafromV}), they have to be considered together rather than separately. Hence, a given  $n$-point function ${\cal G}_{n}$ is allowed only if there exists an (infinite) tower of $n+k$-point functions  ${\cal G}_{n+k}$ which can be reduced to ${\cal G}_{n}$.
One can interpret Eq.(\ref{TfromW0}) just as a particular instance of this reduction, relating the $n$-point function ${\cal G}_{n}$ and the $2n$-point function ${\cal G}_{2n}$. Indeed if ${\cal G}_{2n}$ is known, it is straightforward to compute both ${\cal W}(2n)$ and its restriction ${\cal W}_0(2n)$. 

We have no proof that Eq.(\ref{TfromW0}) is equivalent to the complete set of  relations between the correlation functions of the stress-energy tensor, but there are strong indications that this is indeed the case, since as we shall 
see in the next Section, even a weaker version of it fixes completely the 4-, 5- and 6-point functions of the stress energy tensor.   

As already mentioned, the biharmonic completion (say in $x_1$ and $x_2$) of a rational single pole in all its arguments function is again a rational single pole in $x_1$ and $x_2$ function \cite{biharm1}. It is straightforward to prove that it will be single pole also in all the other arguments. Indeed, assume that in the completion there are terms which violate the single pole condition in some other arguments. Any such terms will necessarily be multiplied by some positive integer power of $Z_{12}$. Denote the smallest such power by $\ell$
and consider the term $(Z_{12})^{\ell} R(\{Z_{ij}\})$, where $R$ by assumption
is single pole in $x_1$ and $x_2$, but is not single pole in some of the other arguments. Imposing harmonicity in $x_1$, recalling that for the $2n$-point functions under consideration the wave operator has the representation $\Box_{x_1}=-4 \sum_{2 \leq j<k \leq 2n} Z_{jk} \partial_{1j} \partial_{1k}$ \cite{Rational2},
noting that only the $j=2$ term can lower the power of $Z_{12}$, and looking at the non single pole part of the coefficient of $(Z_{12})^{\ell-1}$, it follows that $R$ should satisfy $\sum_{3 \leq k \leq 2n} Z_{2k} \partial_{1k} \, R(\{Z_{ij}\}) = 0$, which implies $R=0$. 
The same argument can be repeated for the biharmonic completion in $x_3$ and $x_4$, $x_5$ and $x_6$, etc.,
hence ${\cal W}_2(2n)$, defined in  Eq.(\ref{W2}), will be a rational single pole harmonic in all its arguments function. Note also that we can replace in Eq.(\ref{TfromW0}) the function ${\cal W}_0(2n)$ by 
${\cal W}_2(2n)$.

\section{Computing the correlation functions}
\label{sec:computation}

In the previous Section we derived  Eq.(\ref{TfromW0}), which allows to express  the $n$-point function of the stress energy tensor ${\cal G}_{n}$ in terms of a particular restriction ${\cal W}_0(2n)$ of the $2n$-point function ${\cal G}_{2n}$, or equivalently in terms of its  harmonic completion ${\cal W}_2(2n)$. At first sight this may seem rather useless, since ${\cal G}_{2n}$ is unknown. However, even if we do not know  ${\cal G}_{2n}$, we know quite a lot about the  restriction ${\cal W}_0(2n)$ and its harmonic completion  ${\cal W}_2(2n)$.

In particular ${\cal W}_2(2n)$ is a function of $2n$ variables $x_1,x_2,\dots,x_{2n-1},x_{2n}$, with all the following properties: 
\begin{enumerate} 
\item scalar, 
\item symmetric under the permutation of  $x_{2k-1}$ and $x_{2k}$ for any $k=1, \dots, n$,
\item symmetric under the permutation of any two pairs  $(x_{2j-1},x_{2j})$ and $(x_{2k-1},x_{2k})$, 
\item conformal invariant of conformal weight one in all its arguments,
\item harmonic in all its arguments,
\item rational, \ie of the form (\ref{Wrat}),
\item single pole in all its arguments, 
\item maximal order of all the poles (in $Z_{ij} = x_{ij}^2$) less or equal to five, 
\item finite in the limit $x_{2k} \rightarrow x_{2k-1}$ for any $k=1, \dots, n$. 
\end{enumerate}
The properties from 1 to 5 are true by construction, properties  6 and 7 were proven in the previous section. Let us briefly motivate the remaining two. Since 
${\cal W}_2(2n)$ is the harmonic completion of ${\cal W}_0(2n)$, the order 
of its poles will not exceed the order of the poles of ${\cal W}_0(2n)$, which in turn cannot exceed the order of the poles of the function ${\cal G}_{2n}$, 
which should respect the pole bounds (\ref{polebounds}). This proves property 8. Regarding property 9, the function ${\cal W}_2(2n)$ in fact satisfies a stronger condition.
Namely, due to Eq.(\ref{Vzero}), it vanishes in this limit. We have relaxed this
condition and require only finiteness, since in this way we can treat also the case when in the theory there is a scalar field of dimension two $\Phi_2$ (see the discussion before  Eq.(\ref{Vzero})). We used this case as a cross-chesk of our procedure.   

Since all the properties from 1. to 9. are linear,
for any fixed value of $n$ the general function $F$ with all the above properties will be a finite linear combination $F= \sum_k c_k F_k$,  of functions $F_k$ which have all the properties, with  arbitrary coefficients $c_k$. 
Indeed for most of the 
choices of these coefficients the function  $F$ will not correspond to a restriction of any $2n$-point function of the stress-energy tensor ${\cal G}_{2n}$. However the restrictions of all possible $2n$-point functions ${\cal G}_{2n}$ will certainly correspond to some choice of $c_k$, hence also 
all possible $n$-point functions ${\cal G}_{n}$ can be obtained 
by applying the differential operator in the r.h.s of  Eq.(\ref{TfromW0}) to  $F$.

Hence we proceed as follows :
\begin{itemize}
\item Compute the general function $F(x_1,x_2,\dots,x_{2n-1},x_{2n})$ with the properties from 1 to 9.
\item Compute $\widehat F$ by subtracting from $F$ the  $\Phi_2$ contributions\footnote{It is  sufficient to subtract only up to the second derivatives of the functions of 
$\Phi_2$.} (if present). Note that   
$\widehat F$ vanishes in the limit $x_{2k} \rightarrow x_{2k-1}$ for any $k=1, \dots, n$.
\item Compute 
\be
f(x_1,x_3,\dots,x_{2n-1}) =  \prod_{k=1}^{n}{\cal D}_{2k-1} \widehat F(x_1,x_2,\dots,x_{2n-1},x_{2n}) \vert_{\{x_{2k} = x_{2k-1}\}}
\, ,
\label{TfromF}
\ee
\item Impose on $f$ the pole bounds in Eqs.(\ref{polebounds}) and the cluster decomposition (in order to exclude for $n=6$ the products of 3-point functions). 
\end{itemize}

Finally, renaming the arguments we obtain the function
$f(x_1,x_2,\dots,x_{n})$. 
It will indeed contain the three free field $n$-point functions as special cases. The interesting part is the rest, which we call "plausible candidates" for non-trivial $n$-point function of the stress-energy tensor.

We performed  this calculation  for $n=3,4,5,6$.
The result is: in all cases $f$ contains only the three free field $n$-point functions.

In other words, for any $n \leq 6$ the general rational 
$n$-point function of the stress-energy tensor, defined in Eq.(\ref{generalnp}), is a linear combination of the three free field expressions
\be 
 {\cal G}_{n} = c_{\varphi}\langle\Theta\dots\Theta\rangle_{\varphi}
+c_{\psi}\langle\Theta\dots\Theta\rangle_{\psi}
+c_{F}\langle\Theta\dots\Theta\rangle_{F} \, ,
\label{T6}
\ee
where the coefficients $c_{\varphi}$, $c_{\psi}$ and $c_{F}$ 
are the same for all values of $n$.

Since the spectrum of the fields ${\cal O}_R$ and the coefficient functions ${\cal C}^R$,  appearing in the OPE Eq.(\ref{TTOPE}) are determined 
already from the 4-point functions ${\cal G}_4$,  
one may wonder why we have performed the calculation for up to the 6-point functions. The reason is that the 4-point functions, by a double OPE
expansion (in say $x_{12}$ and $x_{34}$) can be reduced to the 2-point functions
of the fields  ${\cal O}_R$, which are unique. On the contrary, in general there 
are several conformal invariant structures for the 3-point functions
$\langle {\cal O}_{R_1} {\cal O}_{R_2} {\cal O}_{R_3} \rangle $. Hence in principle it could happen that the 4-point functions in two different theories are the same, and the difference appears only at the level of higher-point functions. Our 6-point calculation, by performing a triple OPE expansion, implies that all the 3-point functions of all 
the operators which appear in the OPE of the stress-energy tensor are equal 
to the free field ones. Indeed, it is still possible that the difference shows up only for 7- or higher-point functions, but we consider this extremely unlikely, since nothing qualitatively new can happen above six points.
Hence, we conjecture that Eq.(\ref{T6}) holds for all $n$-point functions of the stress-energy tensor.

Note that so far we never used that the stress-energy tensor is unique.
If one requires that $\Theta_{\mu \nu}$ is the only rank two conserved tensor 
in the $\Theta \Theta$  OPE (in the complete theory there  still  may be several 
different), 
it follows that only one of the three coefficients 
$c_{\varphi}$, $c_{\psi}$ and $c_{F}$ in Eq.(\ref{T6}) can be non-vanishing.  
To prove this let us recall that the functions 
of the stress-energy tensor in the theory of free scalar field have maximal poles of order five,  in the theory of free fermion field have maximal poles of order four, while  in the theory of free Maxwell field have maximal poles of order three. 
Hence, in general the $\Theta$ contribution to the OPE $\Theta \Theta$ has the form
\be
: \Theta(x_1) \Theta(x_2): \vert_\Theta =  
\left({ {\cal C}_\varphi \over (x_{12}^2 )^5 } +
 {{\cal C}_\psi \over (x_{12}^2 )^4 } 
+ {{\cal C}_F \over (x_{12}^2 )^3 } \right) \Theta(x_2) \ , 
\label{unique2}
\ee
where we made explicit the leading light-cone singularities of the different terms.
It follows that the 4-point conformal partial wave of the stress-energy tensor
is
\be
{\cal G}_4 \vert_{\Theta}  = 
\left({ {\cal C}_\varphi \over (x_{12}^2 )^5 } +
 {{\cal C}_\psi \over (x_{12}^2 )^4 } 
+ {{\cal C}_F \over (x_{12}^2 )^3 } \right)
\left({ {\cal C}_\varphi \over (x_{34}^2 )^5 } +
 {{\cal C}_\psi \over (x_{34}^2 )^4 } 
+ {{\cal C}_F \over (x_{34}^2 )^3 } \right)
\langle \Theta \Theta \rangle\ .
\label{Tpwave1}
\ee
On the other hand, if the stress-energy tensor is the only rank two conserved tensor, the same partial wave can be computed also  from Eq.(\ref{T6})\footnote{In general the correlation function will contain a linear combination of the 
partial waves of all rank two conserved tensors.} 
\be
{\cal G}_4 \vert_{\Theta} \ = \ 
\left( { c_\varphi\over (x_{12}^2 x_{34}^2)^5 } 
+  {c_\psi \over (x_{12}^2 x_{34}^2)^4 }
+  {c_F \over (x_{12}^2 x_{34}^2)^3 } 
\right) \langle \Theta \Theta \rangle\  \, . 
\label{Tpwave2}
\ee
The above two expressions are compatible only if any
two of the three coefficients $c_{\varphi}$, $c_{\psi}$ and $c_{F}$
vanish (together with the respective ${\cal C}$ functions).
Hence in a theory with a unique tensor, all the correlation functions 
are proportional to the ones in just one of the three free field theories.
In principle, this does not exclude the theories with several copies of the 
same type of fields. However, recently the correlation functions of the stress-energy tensor in any CFT were studied from a quite different perspective in \cite{Zhiboedov:2013opa} and it was argued that the theories of only scalars or only Maxwell fields are necessarily free.

\section{Conclusions. Open problems}
\label{sec:conclusions}

In this paper, under the assumptions listed in the introduction, we studied the correlation functions of the stress energy tensor in a CFT with HS symmetry and derived general constraints on their singularity structure. We then computed explicitly the 4-, 5- and 6-point functions and found that they are given by a linear combination of the three free field expressions Eq.(\ref{T6}). This is a very strong indication that all such theories, and in particular all the GCI theories, reduce just to the free field ones.

There are many open problems left. 
The first and by far most important is to understand to which extent the rationality assumption restricts the class of CFTs with HS symmetry subject to our analysis. This is particularly important, since our search for non-trivial $n$-point functions gave a negative result. 
The second is to extend our results to all the $n$-point functions of the stress-energy tensor. We think  that already the partial results we found are decisive, but it is always preferable to have a general argument. 

Another interesting open problem concerns the odd rank HS conserved currents.
Although it is immediate to define the (analog of) the scalar bi-field $V(x_1,x_2)$  from the OPE of any two equal HS conserved currents, 
it always projects only on the even rank symmetric tensors, hence it is not clear to us how to define the fusion $ J_r \ J_r  \rightarrow J_{\rm odd}$. This leaves open the possibility that in the  $J_{\rm odd}$ sector something 
interesting may happen, but this goes beyond the scope of this paper.

\section*{Acknowledgements}

\noindent
It is a pleasure to thank K-H Rehren, I.T. Todorov and M.Vasiliev  for
numerous discussions. This work was supported in part by MIUR-PRIN contract 2009-KHZKRX-005.
Hospitality and partial financial support from the Galileo Galilei Institute for Theoretical Physics in Florence during the workshop "Higher Spins, Strings and Duality" is gratefully acknowledged.

\end{document}